\documentclass[sigconf, 10pt]{acmart}

\setcopyright{none}
\renewcommand\footnotetextcopyrightpermission[1]{}
\usepackage{array}
\usepackage{hyperref}
\usepackage{multirow}
\usepackage{graphicx}
\usepackage[utf8]{inputenc}
\usepackage[ruled, lined, longend, linesnumbered]{algorithm2e}
\usepackage{xcolor}
\usepackage{amsmath,multicol}

\usepackage{amssymb,textcomp,comment}
\usepackage{makecell}
\usepackage{ulem}

\usepackage{subcaption}

\usepackage{booktabs}
\usepackage{color}
\usepackage{colortbl}

\usepackage{listings}
\usepackage{enumitem}
\usepackage{cleveref}

\usepackage{pifont}

\usepackage{wasysym}

\usepackage{listings}

\newcommand{\sys}{\texttt{LLMS}\xspace}
\newcommand{\service}{\texttt{LLM Service}\xspace}

\newcommand{\mwx}[1]{\textbf{\color{red}{#1}}}

\begin{document}

\title{LLM as a System Service on Mobile Devices}

\author{Wangsong Yin$^{\blacklozenge}$, Mengwei Xu$^{\Diamond}$, Yuanchun Li$^{\bigstar}$, Xuanzhe Liu$^{\blacklozenge}$}
\affiliation {
	\institution{$^\blacklozenge$Key Lab of High Confidence Software Technologies (Peking University), Beijing, China}
	\country{}
}
\affiliation {
	\institution{$^\bigstar$Institute for AI Industry Research (AIR), Tsinghua University, Beijing, China}
	\country{}
}
\affiliation {
	\institution{$^\Diamond$State Key Laboratory of Networking and Switching Technology (BUPT), Beijing, China}
	\country{}
}

\email{yws@stu.pku.edu.cn}
\email{
	mwx@bupt.edu.cn
}
\email{
liyuanchun@air.tsinghua.edu.cn
}
\email{
liuxuanzhe@pku.edu.cn
}

\begin{abstract}


Being more powerful and intrusive into user-device interactions, LLMs are eager for on-device execution to better preserve user privacy.
In this work, we propose a new paradigm of mobile AI: LLM as a system service on mobile devices (LLMaaS). 
Unlike traditional DNNs that execute in a stateless manner, such a system service is stateful: LLMs execution often needs to maintain persistent states (mainly KV cache) across multiple invocations.
To minimize the LLM context switching overhead under tight device memory budget, this work presents \sys, which decouples the memory management of app and LLM contexts with a key idea of fine-grained, chunk-wise, globally-optimized KV cache compression and swapping.
By fully leveraging KV cache's unique characteristics, it proposes three novel techniques:
(1) Tolerance-Aware Compression: it compresses chunks based on their measured accuracy tolerance to compression.
(2) IO-Recompute Pipelined Loading: it introduces recompute to swapping-in for acceleration.
(3) Chunk Lifecycle Management: it optimizes the memory activities of chunks with an ahead-of-time swapping-out and an LCTRU (Least Compression-Tolerable and Recently-Used) queue based eviction.
In evaluations conducted on well-established traces and various edge devices, \sys reduces context switching latency by up to 2 orders of magnitude when compared to competitive baseline solutions.

\end{abstract}
\maketitle

\section{Introduction}

The recent progress of Large Language Models (LLMs) is reshaping the mobile AI landscape.
LLMs can comprehend human language and handle most (if not all) language-based ML tasks with a huge knowledge base, e.g., language translation~\cite{vaswani2023attention, bahdanau2016neural}, Q\&A~\cite{rajpurkar2016squad, devlin2019bert}, and smart reply~\cite{gboard}.
More importantly, it catalyzes novel mobile applications such as UI automation tasks based on user instructions, e.g., ``forward the recent 5 emails from Bob to Alice''~\cite{wen2023autodroid}.
In a nutshell, LLM marks a giant step for mobile devices towards more intelligent and personalized assistive agent~\cite{li2024personal_llm_agents}.

Being more powerful and intrusive into user-device interactions, LLMs are eager for on-device execution to better preserve user privacy.
For instance, mobile UI automation task takes in the screen information (either in view hierarchy~\cite{wang2023enabling, wen2023autodroid} or pixels~\cite{rawles2023android, yan2023gpt4v, hong2023cogagent}), which could contain highly privacy-sensitive information like chatting history, photos, and textual input.
Beyond privacy, on-device LLM also alleviates the huge resource intensity on datacenters and guarantees functionality availability with weak network.

Indeed, there have been tremendous progress made towards on-device LLM in the past year.
On the one hand, more compact and resource-efficient LLMs (e.g., Gemma-2B and Falcon-1B~\cite{gemma_2024, almazrouei2023falcon}) are released, and various compression algorithms (e.g., 4-bit or even 1-bit quantization~\cite{frantar2023gptq, ma2024era}) are proposed to effectively cut down the resource consumption of LLMs.
On the other hand, system- and hardware-level support for LLMs are maturing~\cite{xu2023llmcad, llama.cpp, song2023powerinfer, xu2024survey}.
For instance, Qualcomm claims Snapdragon 8gen3's NPU to be ``meticulously designed with generative AI in mind'', capable of executing LLM at 20 tokens/second~\cite{8gen3}.
Consequently, smartphone vendors like Google~\cite{tflite-llm} are exploring to built-in LLM into their off-the-shelf devices.

\textbf{LLM-as-a-Service} (LLMaaS).
In this work, we propose a new paradigm of mobile AI:
\textit{LLM as a system service on mobile devices (LLMaaS)}.
It indicates that, the mobile OS exposes an LLM and its inference infrastructure as a system feature to mobile apps, akin to the location or notification services.
The interface between apps and \service is based on prompts in nature language.
This paradigm fundamentally differs from prior arts that apps own their models separately, into which the OS has no visibility.
Such a paradigm shift is natural in LLM era, motivated by following observations:
(1) LLM has world knowledge and can support generic ML tasks~\cite{brown2020language, touvron2023llama, openai2023gpt4, hendrycks2021measuring} through properly curated or even learned prompts.
(2) LLMaaS needs only one copy of LLM weights in memory, regardless of how intensively the LLM is used across apps; otherwise, the LLMs owned by different apps easily blow up the device memory;
(3) A system-level LLM can be better customized for on-device accelerator and enjoy the performance gain over commodity hardware.
An exemplification of LLMaaS paradigm is the recently released Android AICore~\cite{AICore}, a standalone LLM system service that is already in use by several Google apps for on-device summarization and Gboard smart reply.

To fulfill the vision of LLMaaS, this work identifies and tackles a unique system challenge: \textit{LLM context management.}
Specifically, unlike traditional DNNs that execute in a stateless manner, LLMs execution often needs to maintain persistent states (mainly KV cache~\cite{pope2022efficiently}) across multiple invocations.
For example, a smart reply app~\cite{gboard} needs to remember its historical conversation text to generate more accurate reply suggestions compared to using only the last message.
According to our preliminary experiments in $\S$\ref{sec:bkgnd}, a single LLM context could consume significant device memory (e.g., 2GB for Llama2-7B with 4k context window size); more of such LLM contexts soon dominate the memory usage of \service.
Consequently, how to properly manage the persistent LLM contexts across apps becomes crucial to improve the quality-of-LLM-service.
One might treat the LLM context memory as part of the app memory and reuse the mobile memory manager (e.g., low memory killer~\cite{LMK}) to manage them in a unified manner.
This approach, however, is inefficient due to the unique characteristics of LLM contexts as will be demonstrated in $\S$\ref{sec:bkgnd:context}.

\textbf{\sys}
This work presents a first-of-its-kind system towards LLMaaS on mobile devices, named \sys, that decouples the memory management of LLM contexts from the app.
\sys aims to minimize the LLM context switching overhead under tight memory budget, akin to the traditional mobile memory mechanisms that focus on reducing the app cold-start latency~\cite{predictive, Predicting, Practical}.
To alleviate the limited device memory, \sys introduces novel techniques on fine-grained, chunk-wise, globally-optimized KV cache compression and swapping.
\sys splits KV cache into series of chunks -- memory blocks covering the same number of tokens (all layers in one token).
Each chunk is compressed and swapped out/in independently.
In practice, the chunk size is determined empirically based on the configurations of system and LLM, e.g., 16 tokens.
Similar to the OS's paging mechanism, we observe that the idea of chunking strikes good balance between the utilization of device memory and I/O bandwidth, and outperforms token-level or context-level management.

Chunking fully leverages the unique characteristics of KV cache for optimizing context switching.
(1) KV cache is easy to be chunked.
As shown in $\S$\ref{subsec: overview}, its layout grows at the token dimension that can be divided into chunks with flexible granularity.
(2) KV cache is unevenly tolerant to compression.
A portion of KV cache can be compressed more aggressively. 
Compressing in chunk-level allows for maximizing the compression potential of KV cache.
(3) KV cache can be recomputed.
As an intermediate activation of LLM, KV cache can be recomputed from the prompt text to recover it into memory.
By managing KV cache in chunks, during context switching, chunks that need to be loaded from disk can be concurrently recovered into memory through both recompute and I/O, thereby fully utilizing hardware.

To fully explore the design space of chunk-level memory management, \sys incorporates three novel techniques.

(1) Tolerance-Aware Compression ($\S$\ref{subsec: tec_compression}).
\sys uses the information density of a chunk as a metric to quantify its tolerance to compression. 
It calculates the information density based on attention scores, which indicate the level of attention that the tokens pay to each other.
The rationale is that a token that attracts more ``attention'' from other tokens is more likely to be informative.
\sys then judiciously determines the compression rate for each chunk to maximize the overall information intensity of a context, while meeting a global average compression ratio configured by the OS.

(2) Swapping-Recompute Pipeline ($\S$\ref{subsec: ppline}).
\sys recomputes chunks from their original text and overlaps the computation time with the I/O time of other chunks through pipeline.
However, chunks can be swapped independently, while the LLM's continuous position encoding and causal mask cannot handle recompute of interleaved chunks.
Thereby, \sys devises the encoding/mask to fit the interleaved chunks on the fly.

(3) Chunk Lifecyle Management ($\S$\ref{subsec: workingset managing}).
\sys designs the lifecycle of KV cache chunks to be more friendly to context switching.
Regarding which chunk to swap out, it employs an LCTRU Least Compression-Tolerable and Recently Used) queue to determine the eviction priority.
Its rationale is that swapping heavy and least recent used chunks out to disk can better leverage chunks' time locality and \sys's swapping-recompute pipeline.
Regarding when to swap out, it adopts an ahead-of-time swapping-out approach to hide the time for reclaiming memory during context switching.

\textbf{Results }
We have fully implemented \sys on Commercial Off-The-Shelf (COTS) devices including Jetson Orin NX~\cite{orin}, Jetson TX2~\cite{tx2}, and MI14 smartphone~\cite{mi14} with Llama2-7B~\cite{touvron2023llama} and OPT-7B~\cite{zhang2022opt}.
We then evaluate its performance with a 72-hours-long context switching traces synthesized from 6 representative datasets.
The results show that \sys significantly outperforms its baselines. 
It reduces switching latency by up to 2 orders of magnitude compared to managing contexts via the conventional app-level memory manager low-memory killer~\cite{LMK}, or directly managing contexts via disk swapping.
Compared to the state-of-the-art chunk-based context managing system vLLM~\cite{kwon2023efficient} with statically quantizing all chunks to 8-bits~\cite{xiao2023smoothquant}, \sys achieves up to 20$\times$ and on average 9.7$\times$ switching latency reduction.
\sys achieves the aforementioned latency reduction without any noticeable accuracy loss on 6 datasets.
For the first time, \sys addresses the issue of LLM context switching, enabling LLMaaS to provide low switching-latency and stateful services for mobile apps.

\textbf{Contributions}
This work makes following contributions.
\begin{itemize}[leftmargin=10pt,topsep=0pt]
\item We paint a picture of LLM as a system service (LLMaaS) to fully leash the power of LLM on devices, presenting its strong motivations as well as the key challenge of LLM context management in memory.
\item We flesh out LLMaaS with \sys, a concrete \service design based on fine-grained, chunk-wise KV cache compression and swapping.
\sys aims to maximize the context switching speed through a set of novel techniques, including tolerance-aware compression, swapping-recompute pipeline, and strategic chunk lifecycle management.
\item We prototype \sys and comprehensively evaluate its performance on COTS mobile devices and typical LLMs.
The results demonstrate the efficacy of \sys compared to competitive baselines.
\end{itemize}
\section{LLM-as-a-Service: Motivations and Challenges}\label{sec:bkgnd}



\subsection{On-Device Large Language Model}
\begin{figure}[t]
    \centering
      \begin{minipage}[b]{0.28\textwidth}
        \includegraphics[width=\textwidth]{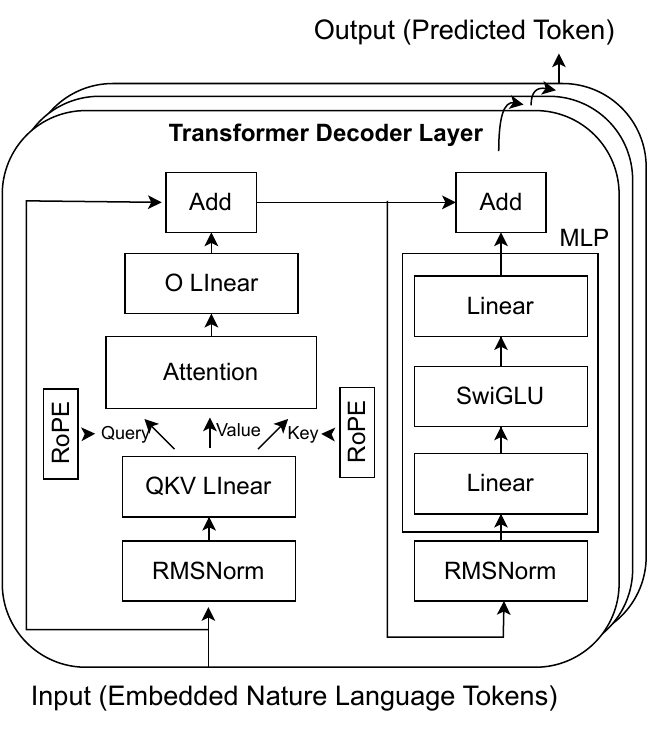}
        \vspace{-17pt}
        \subcaption{\footnotesize The architecture of an LLM. 
        We use Llama2 as an illustrative example.
        Embedding layer is omitted for simplicity.}
        \label{fig: llm-inference}
      \end{minipage}
      \begin{minipage}[b]{0.19\textwidth}
        \includegraphics[width=\textwidth]{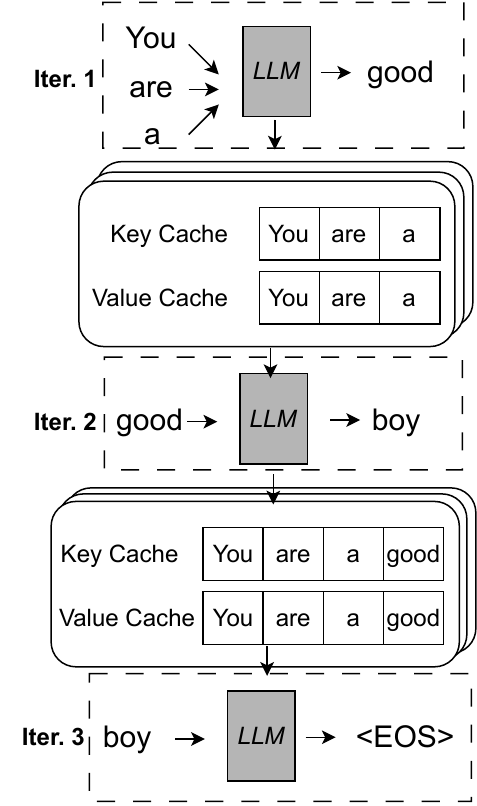}
        \vspace{-17pt}
        \subcaption{\footnotesize 
The autoregressive inference with KV cache.}
        \label{fig:kv cache}
      \end{minipage}
    \vspace{-17pt}
    \caption{Illustrations for LLM’s representative architecture and inference procedure.}
    \label{fig:LLM-inference}
    \vspace{-12pt}
\end{figure}
\textbf{Large language models.}
Large Language Models (LLMs), such as GPT4~\cite{openai2023gpt4}, Llama2~\cite{touvron2023llama}, Gemini~\cite{geminiteam2023gemini}, etc., are transforming mobile AI.
Many cutting-edge applications are empowered by LLM, encompassing agent-based UI automation~\cite{wen2024droidbotgpt, 10.1145/3544548.3580895}, app built-in chatbot~\cite{chatbot}, smart voice assistant~\cite{LLM_based_AI_Assistant}, 
automated email writer~\cite{email}, etc.

We briefly introduce the model architecture and inference procedure of LLMs.
The main body of LLMs is a series of stacked transformer~\cite{vaswani2023attention} decoder layers shown in Figure~\ref{fig: llm-inference}, where input tokens (natural language word pieces) are processed and new output tokens are predicted.
Its key component is the attention mechanism, where tokens are mapped to Query, Key and Value tensors to compute the cross-token connections.
LLMs perform inference in an autoregressive manner: 
in each iteration, it predicts a new token by history tokens (i.e., prompted tokens and generated tokens).
Specifically, LLM caches the Key and Value tensor of previous tokens, known as \texttt{KV Cache}~\cite{pope2022efficiently}, to avoid repeated computations.
We show such an inference procedure in Figure~\ref{fig:kv cache}. 
In iteration 1, the LLM is prompted by ``You are a''.
The model predicts a new token ``good'', and saves the KV cache ``You are a''.
In iteration 2, a new token ``boy'' is jointly predicted by input token `good'' and KV cache ``You are a''.
Meanwhile, the KV cache ``good'' is also saved.
In iteration 3, an ``<EOS>''  token is predicted by input token `boy'' and KV cache ``You are a good'', and the LLM inference ends.


\noindent \textbf{On-device LLM.}
A growing demand is to deploy LLMs on devices to support privacy-preserving mobile AI.
Taking agent-based UI automation as an example,
it takes screen UI as input~\cite{wen2023autodroid}, which is extremely privacy-sensitive as it might include chat history, photos and account information rendered during UI operation.
On-device LLM alleviates such privacy concerns as no data leaves devices.
Moreover, on-device LLM improves service availability and cost efficiency without relying on network and expensive cloud GPUs.

Recently, remarkable progress has been made towards the fast and energy-efficient on-device inference of LLMs, encompassing a full-stack optimization from algorithm to hardware~\cite{8gen3, xu2023llmcad, llama.cpp, song2023powerinfer, xu2024survey}.
For example, mobile SoC vendors have already provided off-the-shelf hardware support for LLMs.
For instance, Qualcomm reports Snapdragon 8gen3's NPU (an ASIC specialized for DNN inference) to be ``meticulously designed with generative AI in mind'', capable of executing LLM at 20 tokens/second~\cite{8gen3}.

\subsection{On-Device LLM as a Mobile OS Service}

We envision a major paradigm shift of mobile AI when on-device LLM matures: \textit{LLM as a system service on mobile devices (LLMaaS)}.
It indicates that, the mobile OS exposes an LLM (as well as the inference infrastructure) as a system feature to apps for use, just like location and notification services, instead of each app owning an LLM individually.
This vision is supported by a recent survey on mobile agent where many industry experts explicitly call for OS-level LLM~\cite{li2024personal_llm_agents}.
Meanwhile, Google has recently released a preview of such LLMaaS design, named AICore, as an Android system service~\cite{AICore}.
The service is already in use by several Google products, such as voice recorder and smart reply.
The interface between apps and \service is text in nature language:
app sends prompts to \service as LLM inputs; \service sends back tokens generated during LLM autoregressive inference.
$\S$\ref{subsec: overview} will show a concrete API design of \service.

Such a vision is backed up by following observations.

\begin{itemize}[left=0pt]
    \item \textbf{LLMaaS is feasible.} LLM has world knowledge and can support generic ML tasks~\cite{brown2020language, touvron2023llama, openai2023gpt4, hendrycks2021measuring}; in-context learning~\cite{dong2023survey, min2022rethinking} and PEFT~\cite{peft, hu2021lora} are capable of further enhancing LLM's capability on downstream tasks.
    A recent study~\cite{yuan2023rethinking} shows that a 7B-sized LLM can achieve better accuracy on most mobile AI tasks compared to the non-LLM models.
    \item \textbf{LLMaaS is desirable.}
    Sharing one LLM across apps guarantees affordable, mostly static memory footprint; otherwise, the LLMs owned by each app easily blow up the device memory.
    The apps might load the LLMs on demand to save memory, yet the weights I/O time easily overwhelms the token generation process: loading LLaMA-7B (4-bit) takes 4.76 seconds, which equals the time to generating 95 tokens on MI 14 smartphone.
    \item \textbf{LLMaaS facilitates hardware and OS design.} On one hand, LLMaaS facilitates the accelerator design on mobile SoCs, since the LLM architecture becomes static and determinable by device vendors.
    A recent paper~\cite{yuan2023rethinking} shows that only the most common 13\% of operators can be fully executed on mobile NPUs.
    On the other hand, LLMaaS grants the full visibility of LLM execution to OS, who thus can schedule, batch, and cache-reuse the inference requests from different apps for better energy efficiency or throughput~\cite{gim2023prompt, orca}.
\end{itemize}

\subsection{LLMaaS Context Managing}\label{sec:bkgnd:context}


A crucial system challenge this work identifies and tackles is \textit{how to efficiently manage the LLM contexts}.
When an app uses \service, it maintains a stateful LLM context, including mainly the KV cache as aforementioned.
Notably, long model contexts are important to LLM-powered mobile apps to provide customized and stateful functionality.
With a long context, the mobile app can define more complex tasks at a time and augment each task with more information so that it can generate more accurate and personalized output.

\begin{figure}[t]
  \begin{minipage}[b]{0.24\textwidth}
    \includegraphics[width=\textwidth]{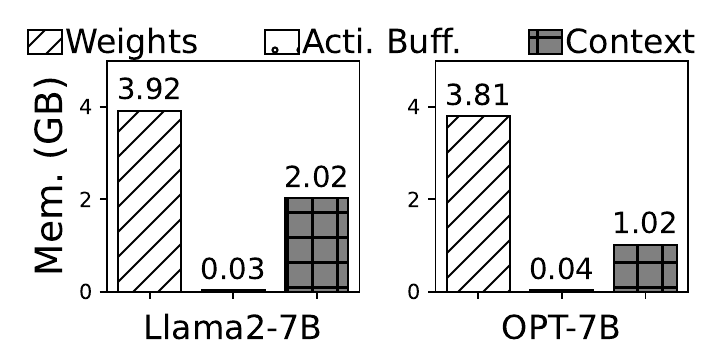}
    \vspace{-17pt}
    \subcaption{Memory breakdown.}
    \label{fig:mem_breakdown}
  \end{minipage}
  \begin{minipage}[b]{0.23\textwidth}
    \includegraphics[width=\textwidth]{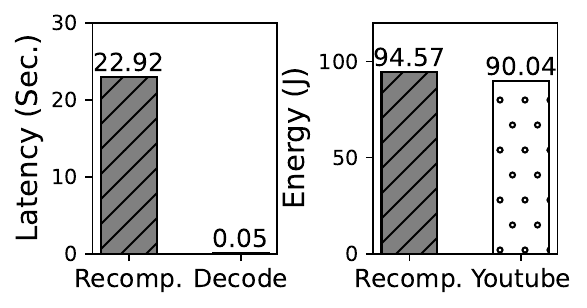}
    \vspace{-17pt}
    \subcaption{Recompute overhead.}
    \label{fig:recompute_overhead}
  \end{minipage}
  \vspace{-10pt}
  \caption{Experiments results on MI14.
  Experiments in (a) is performed on Llama.cpp framework~\cite{llama.cpp}.
  ``recomp.'' in (b) means recompute; ``Decode'' means generating a token; ``Youtube'' means watching videos on YouTube for 1 minute.
  }
  \vspace{-10pt}
\end{figure}

\noindent \textbf{Observation\#1: LLM contexts are memory-intensive.}
Since memory is a scarce resource in executing LLMs~\cite{alizadeh2024llm, yi2023edgemoe, song2023powerinfer}, we study the memory footprint of typical device-affordable LLMs and break down it into three categories: LLM weights, activation buffers, and contexts.
As shown in Figure~\ref{fig:mem_breakdown}, the model contexts contribute significantly to the memory footprint.
For instance, a single context with the maximal context window (4k tokens) of Llama2-7b consumes over 2GB memory, which is over 50\% of model weights.
Note that the LLM weights are shared across all LLM invocations (thereby static), yet the memory usage of LLM contexts proportionally scales with the number of active contexts and longer context window (e.g., up to 128k for recent LLMs~\cite{gpt4turbo}).
Consequently, the LLM context is more likely to be the bottleneck in \service and needs to be carefully managed.

\noindent \textbf{Observation\#2: LLM contexts often need to be persistent.}
In this work, we use the term ``persistence'' to indicate that an LLM context needs to be memorized across multiple invocations that could span hours or even days, regardless of whether apps being switched to background or even killed.
For example, a multi-round conversation agent needs to remember the historical input/output (though with a maximal context window size) whenever it is used, akin to ChatGPT or any other web-based chatting bots.
Besides, an UI automation agent needs to memorize its history operations to serve users' further interactions, such as ``order a pizza in the same restaurant as yesterday''.
Furthermore, the smart reply of Google Gboard calls \service to realize ``generating reply suggestions based on the full context of a conversation, not just a single message.''~\cite{gboard}.
Such persistence feature fundamentally differentiates LLM from prior deep learning models like CNN whose each invocation is independent.

Without system-level persistence support in \service, the apps could instead remember the LLM input/output and feeding them to LLM to ``replay'' (or recompute) the whole LLM context at each invocation.
This approach, however, not only incurs extra programming efforts to developers, but are also highly inefficient on mobile devices.
We conducted preliminary experiments to showcase the resource cost of such context recomputing.
As shown in Figure~\ref{fig:recompute_overhead}, recomputing a Llama2-7B's context takes 22.92 seconds on MI14, while decoding a token for users only takes 0.05 seconds.
The energy consumption of one recompute is roughly equivalent to watching one minute of YouTube video.

\noindent \textbf{Observation\#3: the conventional app-level memory management is not satisfactory.}
One intuitive and plausible design is to account the LLM context memory as part of the whole app's memory who actually uses the LLMaaS, and rely on the app memory manager to manage them as a whole.
For example, mobile devices typically employ low-memory killer (LMK)~\cite{LMK} that kills background apps and frees their memory when system goes out of physical memory.
In this approach, the app memory and LLMaaS context memory are managed together without differentiating them, e.g., both get released if either of them is too large when out of memory.

However, we find this approach not efficient, since the context memory and app memory are inherently different in following aspects.
(1) LLM contexts are more expensive in terms of time/energy expenditure to obtain.
Constructing a context memory takes much time and energy, as this procedure is via LLM inference.
As shown in Figure~\ref{fig:recompute_overhead}, killing an app and recomputing its KV cache incur huge time and energy overhead (e.g., 22.92 seconds and 94.57 J).
(2) LLM contexts are relatively cold, e.g., used in low frequency, even when LLM becomes an indispensable feature on mobile devices.
For instance, Gboard only invokes \service for smart reply functionality during chatting on telegram.
Consequently, a lightweight app could easily gets killed by OS's LMK if it possess an active LLM context, which though will not be used in a near future.
(3) LLM contexts are naturally compressible, as will be discussed in $\S$\ref{subsec: overview}, while app memory are mostly not.
Treating the memory of LLM contexts and app as a whole misses such optimization opportunity.
Instead, we propose decoupling the management of app and model context memory for better LLM efficiency.




\section{\sys design}

\subsection{\sys Overview.}
\label{subsec: overview}

This work advances the vision of LLMaaS with \sys, a first-of-its-kind \service design on devices with a dedicated memory management mechanism of LLM contexts.



\begin{figure*}[t]
    \centering
    \begin{minipage}[b]{0.53\textwidth}
        \includegraphics[width=\textwidth]{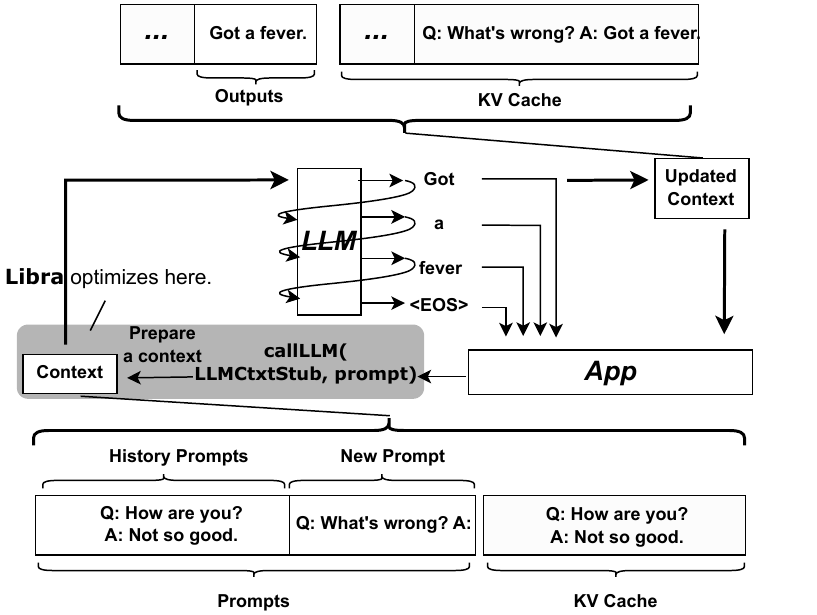}
    \end{minipage}
    \begin{minipage}[b]{0.44\textwidth}
        \input{listing-LLMaaS}
    \end{minipage}

    \vspace{-12pt}
    \caption{Left: a workflow of a chatbot app calling \service via \sys. Right: pseudo codes in Java. 
    } 
    \label{fig:LLMService_wf}
\end{figure*}




\begin{table}[]
\footnotesize
\begin{tabular}{l|l}
\hline
\textbf{Interface} &
  \textbf{Descriptions} \\ \hline
\textit{Class} \texttt{LLMService} &
  \begin{tabular}[c]{@{}l@{}}A system service class that \\ is similar to the Android's \\ \texttt{android.app.Service} class. \end{tabular} \\ \hline
\textit{Class} \texttt{LLMCtx} &
  \begin{tabular}[c]{@{}l@{}}A class that defines LLM context. \\ \texttt{LLMService} interacts with apps \\ via \texttt{LLMCtx}.\end{tabular} \\ \hline
\begin{tabular}[c]{@{}l@{}}\textit{Method} \texttt{newLLMCtx}(\\  ~~~~\textit{Optional} \texttt{systemPrompt}\\ )->\texttt{LLMCtxStub}\end{tabular} &
  \begin{tabular}[c]{@{}l@{}}A method that returns a stub of \\ a new LLM context. On initializing, \\ optional system prompts can be \\ assigned to \texttt{LLMCtx}.\end{tabular} \\ \hline
\begin{tabular}[c]{@{}l@{}}\textit{Method} \texttt{bindLLMService}(\\   ~~~~\texttt{app}\\ )->\texttt{LLMService}\end{tabular} &
  \begin{tabular}[c]{@{}l@{}}A method that binds the \\ \texttt{LLMService} to an app.\end{tabular} \\ \hline
\begin{tabular}[c]{@{}l@{}}\textit{Method} \texttt{callLLM}(\\   ~~~~\texttt{LLMCtxStub}, \texttt{newPrompt}\\ )->\texttt{outputs}\end{tabular} &
  \begin{tabular}[c]{@{}l@{}}A method that calls \texttt{LLMService} \\ via an \texttt{LLMCtx}. It takes an \\ \texttt{LLMCtxStub} and a new prompt as\\ input, and returns the updated \\ \texttt{LLMCtxStub} and decoding results.\end{tabular} \\ \hline
\begin{tabular}[c]{@{}l@{}}\textit{Method} \texttt{delLLMCtx}(\\   ~~~~\texttt{LLMCtxStub}\\ ) \end{tabular} &
  \begin{tabular}[c]{@{}l@{}}A method that deletes an \texttt{LLMCtxStub}. \end{tabular} \\ \hline
\end{tabular}
\caption{\sys interfaces and descriptions.}
\label{tab: interfaces}
\end{table}

\noindent \textbf{\sys APIs.}
As shown in Table~\ref{tab: interfaces}, we design \sys's interfaces to be compatible with Android Services.
Apps interact with \texttt{LLMService} via \texttt{LLMCtx}.
Specifically, we show a chatbot app in Figure~\ref{fig:LLMService_wf}.
Each round, an app appends new prompts to \texttt{LLMCtx} and invokes \texttt{LLMService}. 
When token generation completes, \texttt{LLMCtx} is updated.
Such a chatbot can hold multiple contexts, which are persistent until the app explicitly deletes them through \texttt{delLLMCtx()}.
\sys globally configures the maximal context numbers per app and the maximal context length.
\sys allows each app to hold up to \texttt{K} (a configurable system parameter) active LLM contexts.
Note that each active LLM context has a maximal memory usage with a maximal context window size supported by the LLM.



\noindent \textbf{\sys's design goal.}
As a system service, \sys optimizes for the Quality-of-Service (QoS) it exposes to apps.
Specifically, \sys's design centers around the memory inefficiency challenge as presented in $\S$\ref{sec:bkgnd:context}, aiming to minimize the \textit{context switching latency}, akin to how the mobile memory managers minimize the app cold start latency~\cite{predictive, Predicting, Practical}.
For example, in Figure~\ref{fig:LLMService_wf}, \sys ensures fast context preparation (i.e., make it in memory for LLM inference) when a context is invoked.
This goal has not been explored in prior literature, and is fundamentally different with (as well as orthogonal to) LLM inference speed optimizations during continuous token generation~\cite{song2023powerinfer, xu2023llmcad}.

\noindent \textbf{\sys's context memory model.} \sys exploits a unique design space:
unlike the conventional app memory, LLM context memory is essentially \textit{data chunks that can be approximated.} 
\begin{figure}[t]
    \centering
    \includegraphics[width=0.45\textwidth]{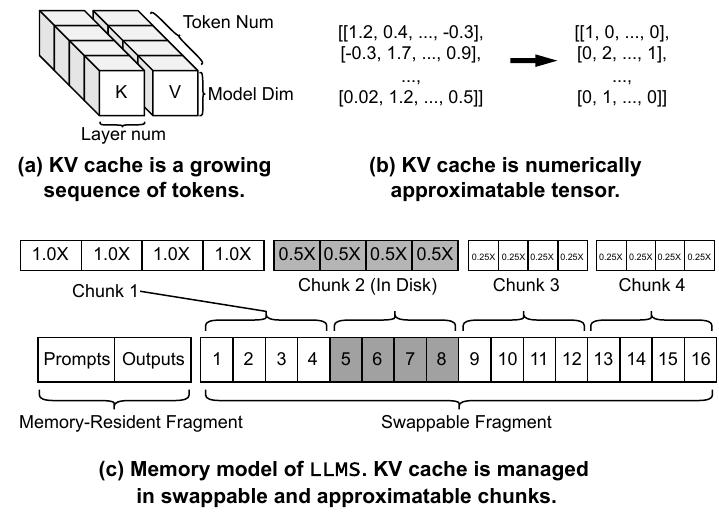}
    \vspace{-15pt}
    \caption{\sys's chunk-wise memory model. Chunks are memory blocks that contain same number of compressed tokens and can be swapped to disk.}
    \label{fig:chunks}
\end{figure}
We show the main body of context memory, KV cache's memory details in Figure~\ref{fig:chunks}(a/b).
Its layout is a growing token sequence that can be divided into chunks with flexible granularity;
each token is a tensor whose numbers can be approximated (e.g. through quantization~\cite{xiao2023smoothquant} or sparsification~\cite{zhang2023h2o}).
Identifying the above characteristics, \sys further introduces \textit{data swapping} to extend the limited device memory, and builds its memory model with swappable and approximatable chunks.
As shown in Figure~\ref{fig:chunks}(c), a context is divided to swappable fragment (KV cache) and memory-resident fragment (prompt/output texts) by \sys.
\textit{The swappable fragment is managed in chunks. }
Each chunk has the same in-chunk compression ratio and same numbers of tokens.
In doing so, \sys makes full use of the limited device RAM and exposes ample virtual memory to contexts.
By carefully selecting the chunk size, \sys can utilize memory with minimal fragmentation compared to managing the context as a whole; compared to token-level managing, a chunk can make better use of IO bandwidth.
Such a chunk size is set empirically to a default number 16 tokens.
We demonstrate its effectiveness and discuss its selection rationale with experiments in $\S$\ref{sebsec: chunksize}.

\begin{figure}[t]
    \centering
    \includegraphics[width=0.48\textwidth]{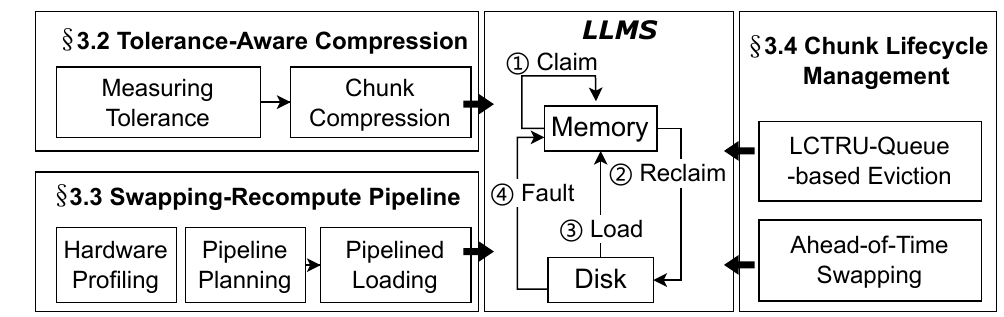}
    \vspace{-17pt}
    \caption{\sys context memory managing overview.}
    \label{fig:sys_overview}
\end{figure}

\noindent \textbf{Overview of \sys's context memory management.}
Figure~\ref{fig:sys_overview} shows an overview of \sys.
The core of \sys manages chunks of context memory during their lifecycle.
It performs the following primitives:
\textcircled{1}\texttt{Claim}, which directly allocates free memory to a chunk;
\textcircled{2}\texttt{Reclaim}, which swaps a chunk out to disk and reallocates its memory to a new chunk when memory is under pressure;
\textcircled{3}\texttt{Load}, which moves a missing chunk from disk to memory before LLM inference;
\textcircled{4}\texttt{Fault}, which moves a missing chunk from disk to memory at each LLM inference iteration. 
In the rest of the section, we introduce the three key techniques of
\sys to meet its design goal:
(1) Tolerance-Aware Compression ($\S$\ref{subsec: compression}) adopts chunk-wise compression to minimize I/O overhead without noticeable accuracy loss;
(2) Swapping-Recompute Pipeline ($\S$\ref{subsec: ppline}) further utilizes the idle computing to accelerate context switching;
(3) Chunk Lifecycle Management ($\S$\ref{subsec: workingset managing}) judiciously decides which chunk and when to swap-out to enhance \sys's QoS.

\subsection{Tolerance-Aware Compression}
\label{subsec: tec_compression}



\noindent \textbf{Chunks exhibit different accuracy tolerance 
to compression
.} 
KV cache compression methods such as quantization or sparsity have been studied by many recent literature~\cite{xiao2023smoothquant, zhang2023h2o}.
However, such methods treat the context as a whole and do not make full use of the idea of chunking.
\sys's key idea is that, different chunks do not contribute coequally to LLM inference.
For instance, a context chunk with tokens like ``context management system'' should contain more information than an ``and so on'' chunk, and the latter should show stronger tolerance to compression (i.e., can be further compressed).
Recognizing that, \sys's chunk compression is tolerance-aware: 
it first applies a conservative KV cache compression to all chunks first, and then iteratively compresses those with higher accuracy-loss tolerance aggressively.

\begin{figure}[t]
    \centering
      \begin{minipage}[b]{0.20\textwidth}
        \includegraphics[width=\textwidth]{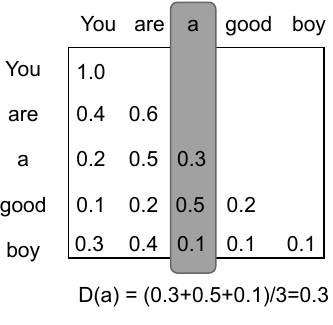}
        \vspace{-17pt}
        \subcaption{\footnotesize Estimating information density via attention scores.}
        \label{fig:attentionscore}
      \end{minipage}
      \begin{minipage}[b]{0.27\textwidth}
        \includegraphics[width=\textwidth]{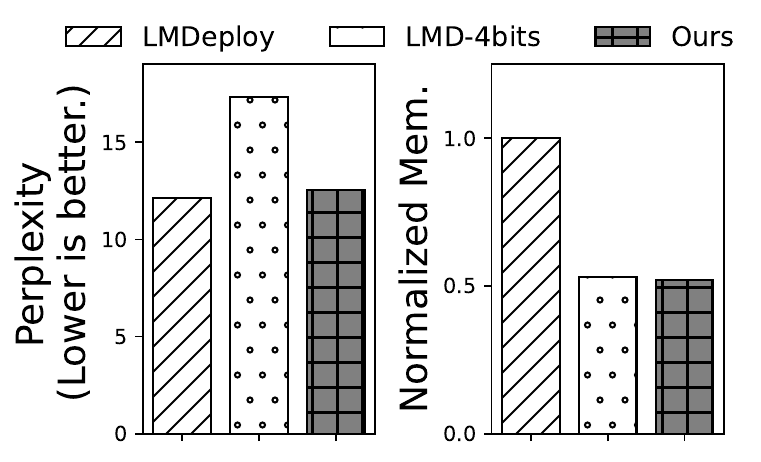}
        \vspace{-17pt}
        \subcaption{\footnotesize Comparisons between LMDeploy's context-level compression and \sys's chunk-level compression.}
        \label{fig:compression_data}
      \end{minipage}
    \caption{\sys's tolerance-aware compression.}
    \label{fig:compression}
\end{figure}

\noindent \textbf{Measuring the tolerance.} Compression tolerance is measured by the information density.
Information density $D_{i}$ of the $i_{th}$ chunk is calculated by 
the attention scores:
\begin{align}
    D_{i} = \frac{1}{q-p}\sum^{q}_{col=p}\frac{1}{L}\sum^{L}_{l=0}\frac{1}{H}\sum^{H}_{h=0}(\frac{1}{R-row}\sum^{R}_{row=0}A^{l, h}_{row, col})),
    \label{eq:attn}
\end{align}
where $A^{l, h}_{row, col}$ is the attention score of the $l_{th}$ layer and $h_{th}$ head, and the $i_{th}$ chunk contains tokens from the $p_{th}$ to $q_{th}$.
Specifically, as shown in Figure~\ref{fig:attentionscore}, the attention score is a $R$ rows and $C$ columns lower triangular matrix calculated by $A = softmax(mask(\frac{Q\cdot K^T}{\sqrt{d_{k}}}))$~\cite{vaswani2023attention}.
Each row represents the ``attention score'' that a token pays to others.
The scores are softmaxed and sums to 1.0 at each row.
For instance, the number ``0.5'' at ``a'' row and ``are'' col means that the token ``a'' pays 50\% attention to ``are''.
If a token is always paid more attention by other tokens, it should be more informative.
Thereby, \sys estimates a token's information density by averaging its column in attention score matrix.
In Figure~\ref{fig:attentionscore}, the information density of token ``a'' is $(0.3+0.5+0.1)/3=0.3$.
As shown in Equation~\ref{eq:attn}, such token-level information density is further accumulated by heads, layers and tokens to achieve chunk-level density, i.e., compression tolerance.

\noindent \textbf{Compressing chunks.}
\sys provides multiple levels of compression ratio for chunks, denoted as $\{ratio_{w}\}$.
Before compression, \sys computes each chunk's $D_{i}$ and determines its ranking $Rank_{i}$(\%) among all other chunks in the context.
Then a series of thresholds $\{\sigma_{ratio}\}$ are formed.
Chunk $i$ is compressed to $ratio_{w}$ based on $Rank_{i}$, \textit{s.t.},
\begin{align}
    \sigma_{ratio_{w+1}} < Rank_{i} \leq \sigma_{ratio_{w}}.
    \label{eq:rank}
\end{align}
Specifically, the thresholds are formed by maximizing the overall information intensity of a context under a given global average compression ratio $ratio_{global}$, which is configurable by the OS.
\sys maximizes
\begin{equation}
\begin{split}
    ctxInfo = \sum_{w}\frac{1}{ratio_{w}}\sum_{\sigma_{ratio_{w+1}} < Rank_{i} \leq \sigma_{ratio_{w}}}D_{i}, \\
    s.t., \sum_{w}ratio_{w}\cdot(\sigma_{ratio_{w}}-\sigma_{ratio_{w+1}}) = ratio_{global}.
    \label{eq:thresh}
\end{split}
\end{equation}

In practice, \sys adopts quantization for compression. 
During LLM inference, the generated KV cache is quantized by state-of-the-art context-level quantization methods~\cite{xiao2023smoothquant, LMDeploy} that have already been built in LLM.
\sys further performs channel-wise linear~\cite{han2016deep} quantization with lower bitwidth.
For instance, atop an 8-bits quantization method, \sys can further provide 4-bits and 2-bits quantization for some chunks.
Formula~\ref{eq:thresh} can be optimized through a simple differentiation in this case, as there is only one variable.
Note that \sys's tolerance-aware compression is general.
When the LLM's default KV quantization algorithm is 4-bits, it can still work by providing 2-bits and 1-bit further quantization.

\noindent \textbf{Micro experiments} are conducted to show the effectiveness of \sys's tolerance-aware compression.
We adopt the KV cache quantization method of a state-of-the-art LLM inference framework LMDeploy~\cite{LMDeploy}, whose default quantization bitwidth is 8-bit (INT8).
We set \sys compression ratios to $\{ratio_w\} = \{8/8, 4/8, 2/8\}$.
With the global compression ratio $ratio_{global}$ set to 50\%, we run language modeling with Llama2-7B model on WikiText-2 dataset~\cite{merity2016pointer}.
As shown in Figure~\ref{fig:compression_data}, we compare our method to LMDeploy's 8-bits (LMDeploy) and 4-bits (LMD-4bits) quantization.
Our method achieves comparable accuracy (perplexity) to 8-bits quantization and comparable memory consumption compared to 4-bits quantization.
We show detailed discussion of our method's efficacy and rationale of ratio selection in $\S$\ref{subsec: compression}.


\subsection{Swapping-Recompute Pipeline}
\label{subsec: ppline}

The Load primitive in \sys loads missing chunks from disk to memory upon calling \texttt{callLLM()}.
\sys introduces recompute to swapping-in to further accelerate it, as processor is idle during disk I/O.
Recompute here means that use the original prompt text peices to calculate a portion of KV cache, as KV cache is essentially LLM activations.
By recomputing some chunks in a pipelined maner instead of loading them through disk I/O, we can fully utilize the hardware. 

\begin{figure}[t]
    \centering
    \includegraphics[width=0.47\textwidth]{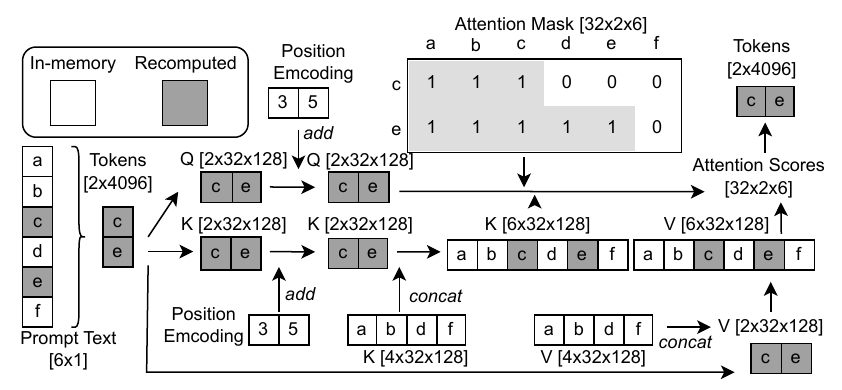}
    \caption{An example of \sys's chunk-recomputing procedure. We use a Llama2-7B layer here for representation. It has 32 heads and 4096 hidden size. \sys recomputes the missing ``c'' and ``e'' in KV cache based on their prompt text and other tokens' KV cache.}
    \label{fig:pipeline}
\end{figure}

\noindent \textbf{Making interleaved chunks recomputable.}
Recall that in \sys's context memory model, each chunk can be swapped out to disk.
Thereby, \sys faces a challenge: employing recompute to recover interleaved non-contiguous chunks.
To do so, \sys refines the LLM recompute procedure.
Given a chunk-wise KV cache $\mathbf{C} = \{chunk_i\}$ and corresponding prompt text $\mathcal{T} = \{text_{i}\}$, \sys recovers the missing chunks $\{chunk_o\}$ through $\{text_o\}$ and $\{chunk_i\}-\{chunk_o\}$.
Specifically, Figure~\ref{fig:pipeline} gives an example of \sys's chunk-recomputing procedure.
In Figure~\ref{fig:pipeline}, the KV cache of a text sequence ``a b c d e f'' is partly swapped out of memory (``c'' and ``e'').
To recover, \sys embeds ``c'' and ``e'' to tokens, and recomputes them to get Q, K and V tensors.
Here, \sys applies a global position encoding to Q and K, i.e., encoding ``c'' with position 3 and ``e'' with position 5.
Then, the recomputed K/V of ``c'' and ``e'' is inserted to the K/V of ``a'', ``b'', ``d'' and ``f''; the entire K/V is recovered in current layer.
After that, Q is calculated with the recovered K to get the attention scores. 
The attention mask is a causal mask~\cite{vaswani2023attention} that retains all tokens before the current token.
In Figure~\ref{fig:pipeline}, the attention mask masks out ``d e f'' for ``c'' and ``f''  for ``e''.
Finally, the tokens that input to the next layer are calculated by the attention scores and the recovered V.
In doing so, \sys realizes the exact recompute of any chunks.

\noindent \textbf{Swapping-recompute pipeline.}
\sys concurrently recomputes a portion of chunks and swaps other chunks into memory from disk.
\begin{figure}[t]
    \centering
    \includegraphics[width=0.47\textwidth]{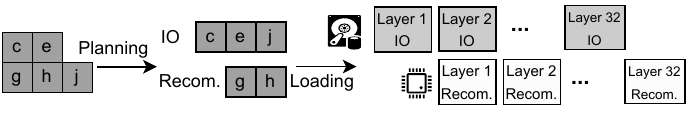}
    \caption{Swapping-recompute pipeline.}
    \label{fig:pipeline_overlap}
\end{figure}
As shown in Figure~\ref{fig:pipeline_overlap}, I/O and recompute are overlapped in a pipeline, where the next layer's I/O is performed during the current layer's recompute.

Such a pipeline is elastic: \sys can adjust the loading configuration (i.e., which chunks are recomputed and which chunks are swapped) to maximize the efficiency.
\sys plans the configuration of each loading based on offline profiled hardware information.

\paragraph{\romannumeral 1. Profiling.}
Recompute delay $T_{re}(x, f, e)$ is a function w.r.t. the number of chunks $x$, the hardware frequency $f$ and the energy mode $e$.
IO delay $T_{IO}(m)$ is a function w.r.t. the size $m$ of on-loading chunks.
In practice, we approximate $T_{re}$ and $T_{IO}$ with linear functions.
\sys performs a cost-effective one-shot measurement with discrete test points at installation time.
The above functions fit these test points.

\paragraph{\romannumeral 2. Planning.}

\sys plans its elastic pipeline by solving the following optimization.
Given the on-loading memory size $m$ and the number of chunks $\{x_{ratio_{w}}\}$ with different compression ratios, \sys minimizes the following equation by determining a proper 
number of recomputed chunks $\{x^{re}_{ratio_{w}}\}$ with different compression ratios.
\begin{equation}
\begin{split}
    pipelineDelay = max[T_{re}(\sum_{w}{x^{re}_{ratio_{w}}}),\\ T_{IO}(m - \sum_{w}{ratio_w \cdot x^{re}_{ratio_{w}}})], \\
    s.t., \forall w, x^{re}_{ratio_{w}} < x_{ratio_{w}}.
    \label{eq:ppl}
\end{split}
\end{equation}
Such a problem is solved by a linear programming, which incurs negligible overhead on devices.

\subsection{Chunk Lifecycle Management}
\label{subsec: workingset managing}




\sys manages the lifecycle of KV cache chunks to be more friendly to context switching.
It mainly decides which chunk and when to swap out.
Regarding which chunk to swap out, it employs an LCTRU queue to determine the eviction priority.
Regarding when to swap out, it adopts an ahead-of-time swapping-out approach to hide the time for reclaiming memory during context switching.

\paragraph{\romannumeral 1. AoT Swapping.}
Compared to complex memory modification timing of apps~\cite{atcswap, mars, SmartSwap}, \service's memory modification is much easier to monitor: chunks are sequentially modified during LLM inference.
Thus, swapping-out can be performed ahead of relaim.
\sys swaps out all the modified chunks at the returning stage of \texttt{callLLM()} (even when not under memory pressure).
Its delay is imperceptible to the \service caller.
In doing so, the reclaim primitive at context preparation stage is overhead-free.

\paragraph{\romannumeral 2. LCTRU-Queue-based Eviction.}
A good eviction policy can reduce the overhead of swapping.
App memory managing methods, such as LMK, evict memory based on the app types (ranked by oom\_adj\_score).
As a system service, \sys does not differentiate the type of context owner.
Its eviction policy is based on two principles: 
\textit{i)} leveraging contexts' time locality;
\textit{ii)} heavy chunks should be evicted first.
Principle\textit{ ii} is derived by the swapping-recompute pipeline in $\S$\ref{subsec: ppline}.
Given memory size $m$, the total number of chunks is inversely proportional to the number of less-compressed chunks.
According to Equation~\ref{eq:ppl}, under the same memory size, a smaller number of chunks will result in a lower pipeline delay, as the recomputing delay $T_{re}(x, f, e)$ is irrelevant to memory size.

Based on the above principles, \sys's employs an LCTRU (Least Compression-Tolerable and Recently-Used) queue. The LCTRU queue consists of multiple concatenated sub-queues with different compression ratios, i.e., $Q_{LCTRU} = \{Q_{ratio_{w}}\}$.
Each sub-queue is ordered by the recently accessed time of its in-memory chunks.
$Q_{LCTRU}$ is updated upon each invocation of \texttt{callLLM()}.
When memory reclaiming occurs, \sys pops out the corresponding number of elements from $Q_{LCTRU}$ based on the required memory size.

Besides, \sys manages a context's chunks as a memory-resident working set:
during \texttt{callLLM()}, \sys \textit{locks} the context memory, forbidding reclaiming their own chunks.
In doing so, \sys avoids system thrashing and realizes being transparent to LLM inference.
Notably, although it will not be triggered by \sys, the Fault primitive is still retained for robustly handling exceptions such as system crush.

\section{Implementation and Methodology}
\label{sec:impl}

\begin{table}[]
\footnotesize
\begin{tabular}{l|l|l|l}
\hline
\textbf{Device Name} & \textbf{\begin{tabular}[c]{@{}l@{}}RAM \end{tabular}} & \textbf{Disk} & \textbf{Hardware Accelerator} \\ \hline
Jetson Orin NX & 8 GB & NVMe SSD     & 1024-core Ampere™ GPU   \\ \hline
Jetson TX2     & 8 GB & SATA HDD     & 256-core Pascal™ GPU   \\ \hline
MI14 Smartphone& 8 GB & UFS 4.0      & Hexagon™ 8Gen3 NPU \\ \hline
\end{tabular}
\caption{Details of mobile/edge devices we use.}
\label{tab:devices}
\end{table}

\textbf{\sys implementation.} We have fully implemented a \sys prototype with 3.5k LoC in Python/C++.
We implement an \service on three representative COTS mobile/edge devices shown in Table~\ref{tab:devices}.
Jetson Orin NX/TX2~\cite{orin, tx2} are high-end edge boards for autonomous robotics or cars.
MI14~\cite{mi14} is a smartphone equipped with UFS4.0 storage and Hexagon 8Gen3 NPU.
We build \service atop Huggingface Transformers~\cite{wolf-etal-2020-transformers} and mllm~\cite{mllm}.
The former is the most popular off-the-shelf LLM deployment framework on devices with Pytorch support~\cite{pytorch}.
On smartphones, we choose the latter because it is lightweight and resource-efficient.
The \service runs as an independent process.
It receives inference requests from client processes through socket IPC.
Except for the client processes and the \service process, we do not run any other non-essential application processes on device.
We select two LLMs: Llama2-7B~\cite{touvron2023llama} and OPT-6.7B~\cite{zhang2022opt}, with a maximal context length as 4K and 2K, respectively.
We apply a sliding window~\cite{xiao2023streamingllm} to the context to enable LLM inference in a streaming manner.
The weights are downloaded from the official repositories on huggingface website.
We quantize LLM weights to 4-bit integers with GPTQ~\cite{frantar2023gptq}.
The KV cache is stored in 8-bits integer by default~\cite{xiao2023smoothquant};
\sys further applies chunk-wise compression to it with its tolerance-aware compression technique ($\S$\ref{subsec: compression}) with a global compression ratio $ratio_{global}$ as 50\%.
We select three levels of chunk-wise compression ratio, i.e., $\{ratio_w\} = \{8/8, 4/8, 2/8\}$. 


The context memory management module of \sys is embedded within \service.
We use pickle~\cite{Pickle} and pickle-in-cpp~\cite{Pickle-in-Cpp} to implement memory-disk swapping.
The chunk size is set to 16 tokens.
Since the sub-byte data format is not natively supported by the LLM inference framework, \sys utilizes parallel bit-shift operations to pack the compressed data into a supported INT8 format.
The swapping-recompute pipeline is implemented by multithread.
We use an independent I/O thread to load chunks from disk to memory.
The computation thread proceeds to the next layer only after the I/O thread for the current layer (reading the next layer's KV cache) has completed.

\begin{table}[]
\footnotesize
\begin{tabular}{l|l|l}
\hline
\textbf{Task}            & \textbf{Dataset} & \textbf{Delta length} \\ \hline
News Classification      & AGnews~\cite{Zhang2015CharacterlevelCN}           & 0.2k--0.5k            \\ \hline
Document Summary         & Xsum~\cite{Narayan2018DontGM}             & 1k--2k                \\ \hline
Chat History Summary     & Samsum~\cite{gliwa-etal-2019-samsum}           & 0.1k--0.3k            \\ \hline
Text Comprehension       & cnn dailymail~\cite{DBLP:conf/nips/HermannKGEKSB15}    & 0.5--1k               \\ \hline
Translation              & WMT17-de-en~\cite{bojar-EtAl:2017:WMT1}      & 0.1k--0.5k            \\ \hline
Sentiment Classification & SST-2~\cite{bojar-EtAl:2017:WMT1}            & 0.01k--0.1k           \\ \hline
\end{tabular}
\caption{Datasets we use for trace synthesis. An entry in a dataset is regarded as one LLM calling. "Delta length" refers to the length of context growth after each \texttt{callLLM()}.}
\label{tab:datasets}
\vspace{-12pt}
\end{table}

\begin{figure*}[t]
    \centering
    \includegraphics[width=0.99\textwidth]{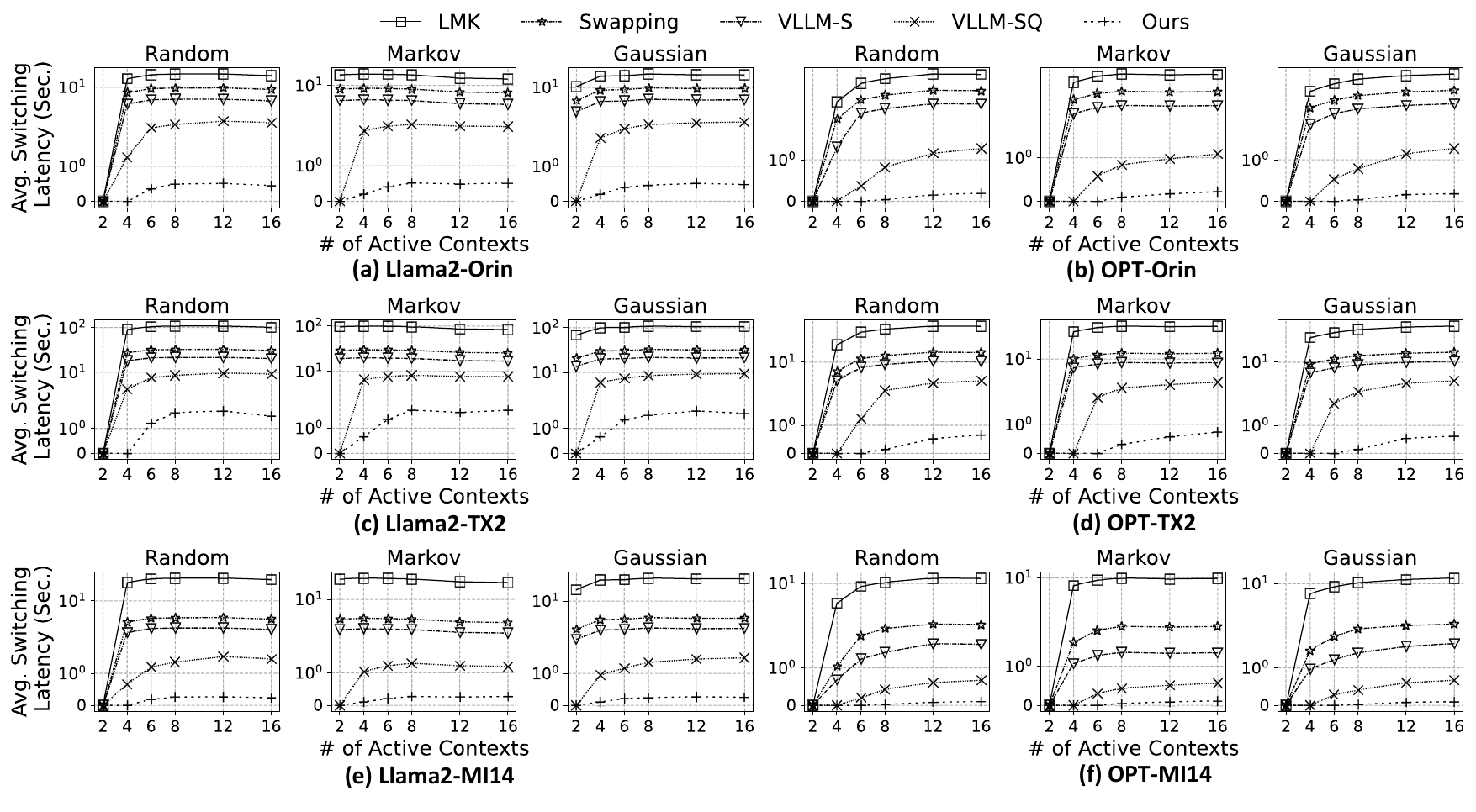}
    \vspace{-17pt}
    \caption{On-average context switching latency on a 72-hours-long trace.}
    \label{fig:mainexp}
\end{figure*}
\begin{figure}[t]
    \centering
    \includegraphics[width=0.48\textwidth]{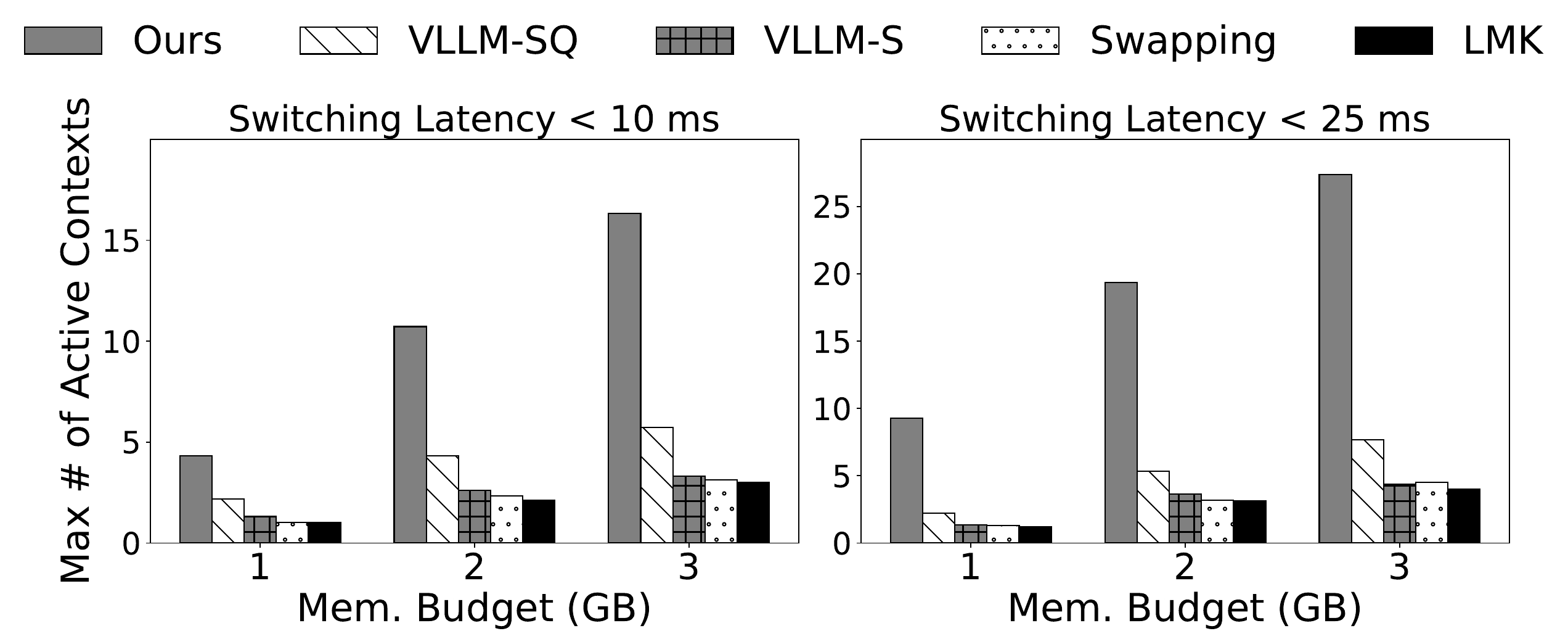}
    \vspace{-17pt}
    \caption{Performance under various memory budgets. Model: Llama2; device: Orin.}
    \label{fig:mem_budget}
    \vspace{-12pt}
\end{figure}

\begin{figure}[t]
    \centering
    \includegraphics[width=0.48\textwidth]{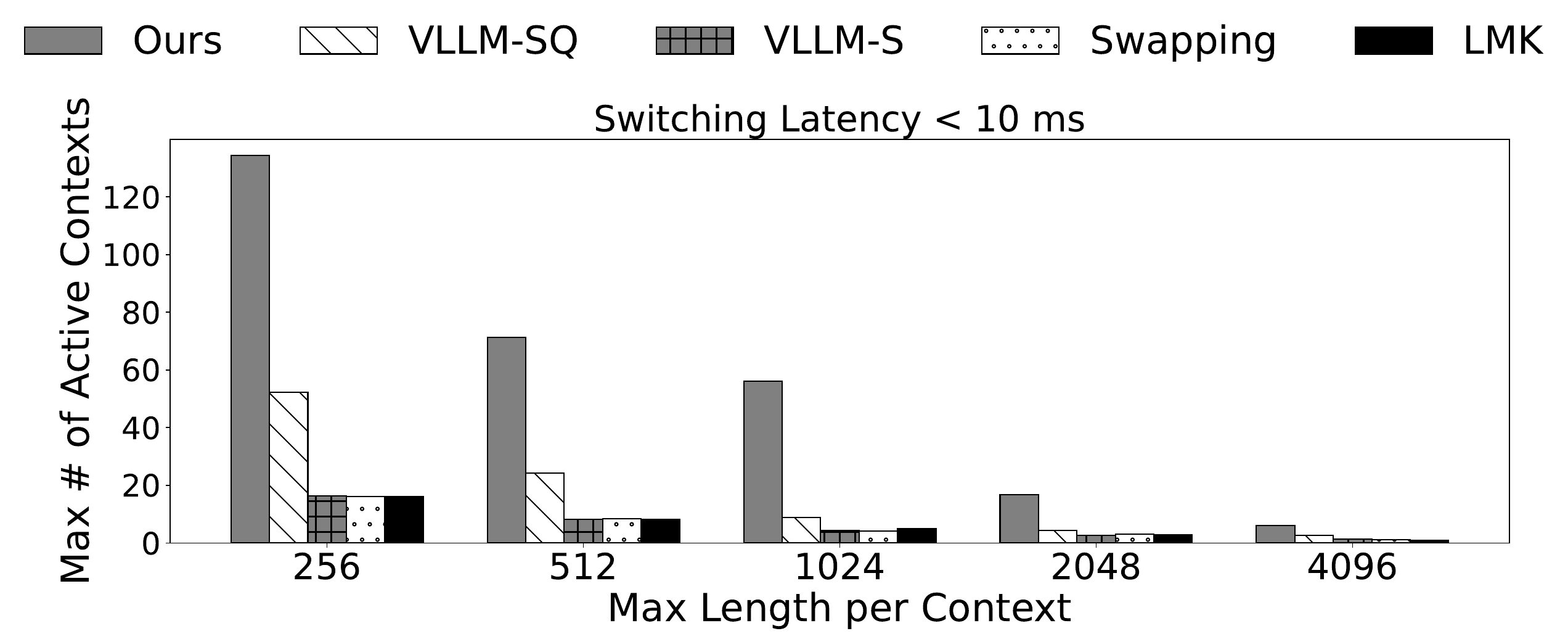}
    \vspace{-12pt}
    \caption{Performance under various maximal context lengths. Model: Llama2; device: Orin.}
    \label{fig:ctxt_length}
    \vspace{-12pt}
\end{figure}


\noindent \textbf{Context switching trace.} To the best of our knowledge, there is no publicly available trace of LLMaaS context switching on devices.
In order to comprehensively and accurately evaluate \sys, we synthesized a trace that formulated by
\begin{equation}
    Trace = \{(Time_{i}, CtxtID_{i}, Prompt_{i}, groundTruth_{i})\},
\end{equation}
Where $Time_{i}$ is the $i_{th}$ calling time of \texttt{callLLM()} with $CtxtID_{i}$, and $Prompt_{i}$ and $groundTruth_{i}$ are the input and ideal output text.
The prompts and groundTruths for a context are derived from a dataset in Table~\ref{tab:datasets}, while a dataset can derive multiple contexts.
We generate calling time $Time_{i}$ using Poisson distribution with different calling rates.
Akin to apps' switching~\cite{predictive, Predicting, Practical}, context switching pattern could be complex: it can be either irregular or with preference (e.g., influenced by invoke history or workloads).
Recognizing that, we construct the following various patterns of context switching to simulate real-world scenarios.

\noindent $\bullet$~\texttt{Random}.
Contexts are switched with the same probability.

\noindent $\bullet$~\texttt{Markov}.
Context switching is determined by a first-order Markov process, which assigns higher priority to recently used contexts.

\noindent $\bullet$~\texttt{Gaussian}.
Context switching follows a Gaussian distribution w.r.t delta length. 
In this pattern, a context with moderate workload is more likely to be requested by the apps.

Note that \sys will not try to predict the context switching pattern, as such a pattern cannot be known as a priori.

We synthesize 72-hours-long traces with different settings to evaluate \sys.
We will make the traces used in our experiments publicly available for reproducibility and further research on on-device LLMaaS.

\noindent \textbf{Baselines.}
We compare \sys to the following alternatives by reporting the performance on the same trace:

\noindent $\bullet$~\texttt{LMK}.
Contexts are killed by a low-memory killer~\cite{LMK} when memory is under pressure.
The killed contexts need to be recomputed when called again.

\noindent $\bullet$~\texttt{Swapping}.
Contexts are swapped out to disk as a whole when memory is under pressure.
The swapped-out contexts need to be swapped-in when called again.

\noindent $\bullet$~\texttt{VLLM-S}.
VLLM~\cite{kwon2023efficient} is a state-of-the-art KV cache managing system without compression.
We reproduce its chunk-wise KV cache managing on devices. 
Chunks are swapped under memory pressure.

\noindent $\bullet$~\texttt{VLLM-SQ}.
KV cache chunks in \texttt{VLLM-S} are further compressed by a state-of-the-art activation quantization algorithm SmoothQuant~\cite{xiao2023smoothquant}.
Chunks are equally quantized to the same level (INT8).

\noindent \textbf{Metrics.}
We mainly report two metrics of \service's context switching QoS:
\textit{Context Switching Latency on Average} and \textit{Maximal Number of Active Contexts}.
The former is under a given number of active contexts; the latter is under a given switching latency constraint.
``Active context'' refers to contexts that are persistent (have not been deleted by \texttt{delLLMCtx()}) and could be invoked.

\section{Evaluation}

\begin{figure}[t]
    \centering
      \begin{minipage}[b]{0.23\textwidth}
        \includegraphics[width=\textwidth]{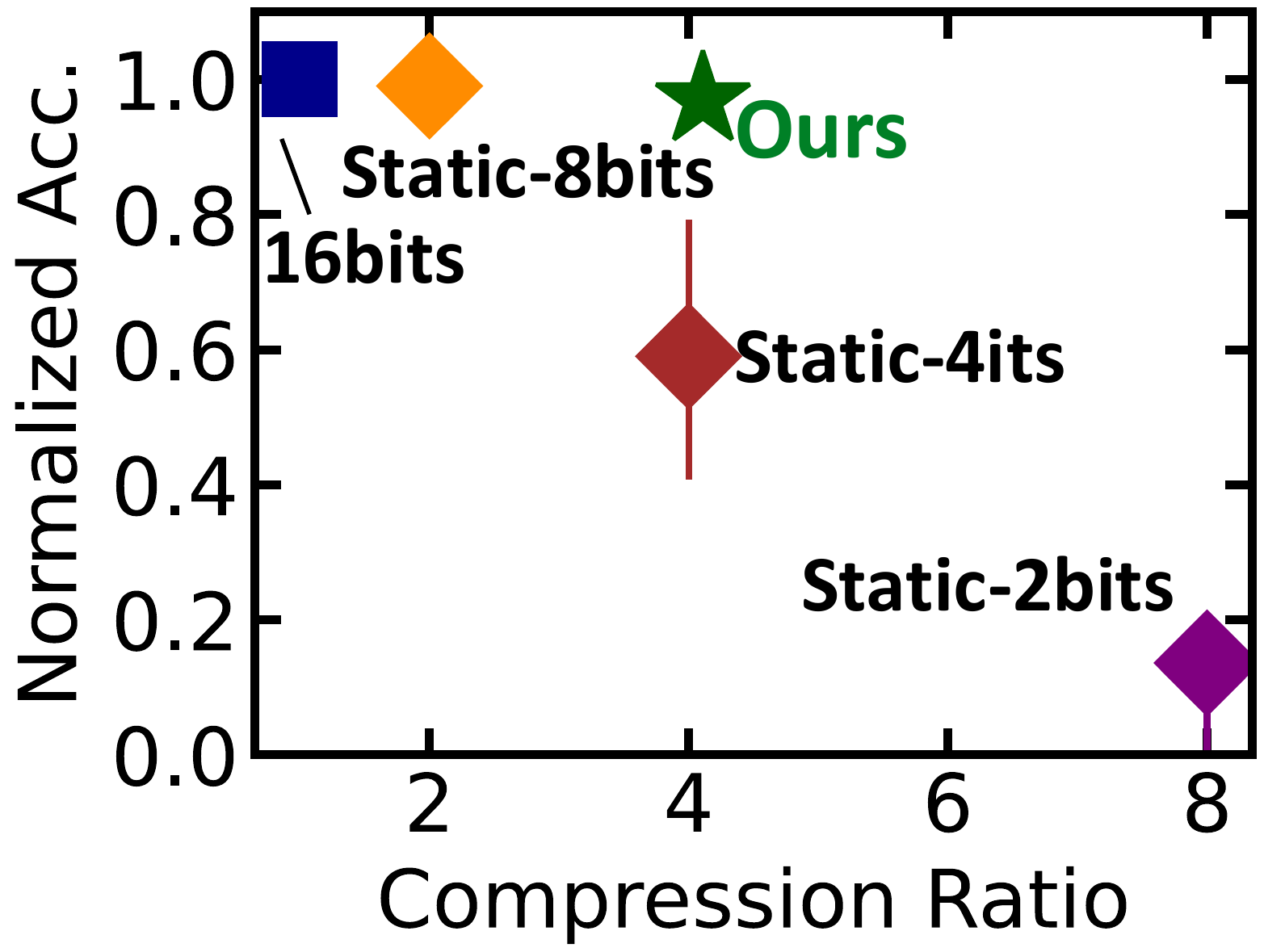}
        \vspace{-17pt}
        \subcaption{Llama2}
      \end{minipage}
      \begin{minipage}[b]{0.23\textwidth}
        \includegraphics[width=\textwidth]{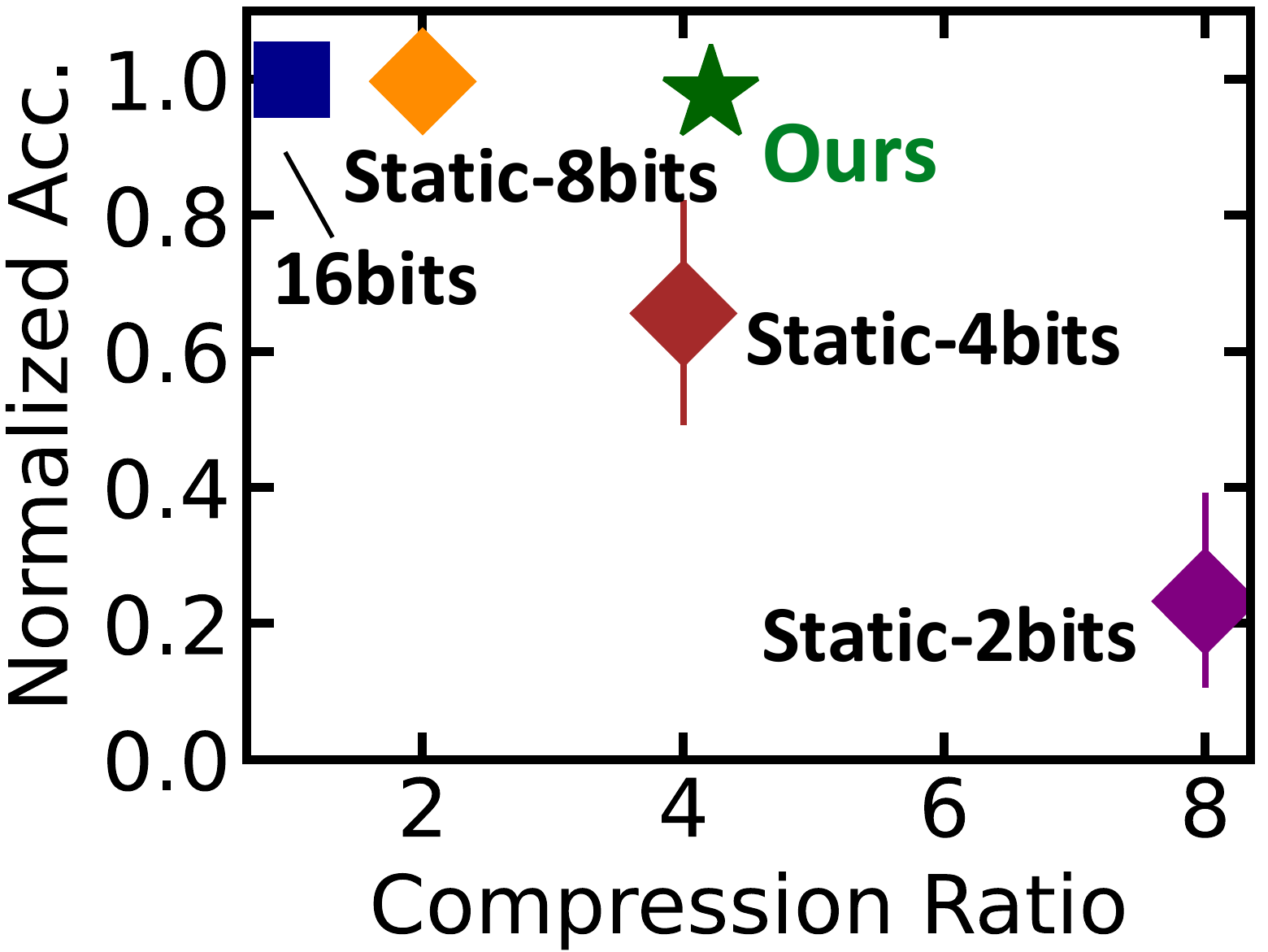}
        \vspace{-17pt}
        \subcaption{OPT}
      \end{minipage}
    \vspace{-14pt}
    \caption{Compression ratio and accuracy loss of static quantization v.s. our quantization. The compression ratio is taken reciprocally here, meaning the higher, the better.}
    \label{fig:quant}
    \vspace{-12pt}
\end{figure}
\begin{figure}[t]
    \centering
    \includegraphics[width=0.46\textwidth]{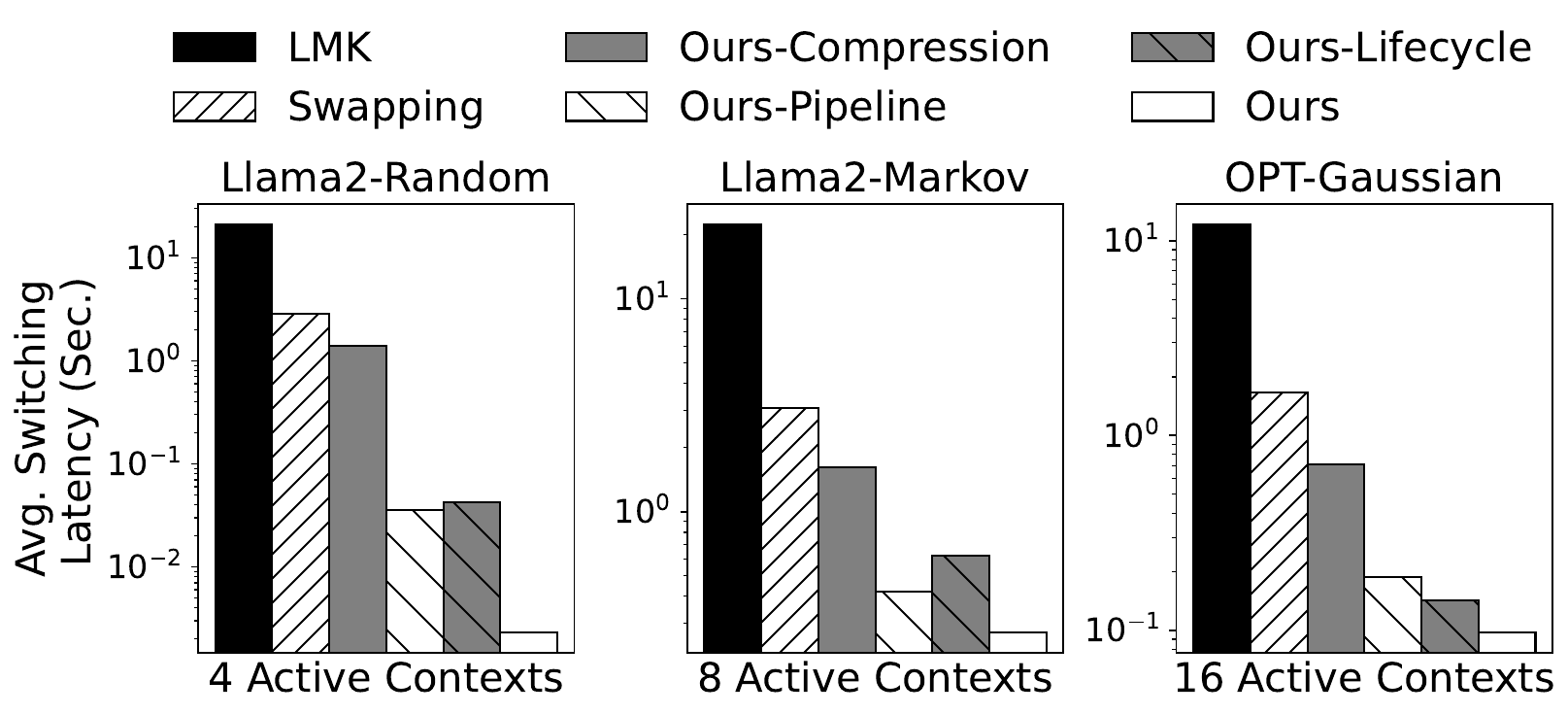}
    \vspace{-14pt}
    \caption{Ablation study.}
    \label{fig:ablation}
    \vspace{-12pt}
\end{figure}

\subsection{End-to-End Context Switching Performance}
\label{subsec:e2e_performance}

We first evaluate to what extent \sys enhances the context switching QoS of \service.

\noindent \textbf{End-to-end switching latency.}
We report the on-average switching latency of 2/4/6/8/12/16 active contexts by running the synthesized trace in $\S$\ref{sec:impl}.
The maximal retained context length is set to the LLM's default context window (4k for Llama2 and 2k for OPT).
The calling rate is one request in five minutes on average.
All remaining memory after running the \service on the device is dedicated to contexts.
The experimental results are shown in Figure~\ref{fig:mainexp}. 
The y-axis has been taken with symmetric logarithmic scale.
We have the following observations. 

In general, compared to the de facto app memory managing method \texttt{LMK}, \sys achieves 
a significant reduction in switching latency by up to 2 orders of magnitude;
compared to the vanilla \texttt{Swapping} baseline, \sys reduces switching latency by 1--2 orders of magnitude;
compared to other strong baselines that applies chunking+swapping (\texttt{VLLM-S}) and chunking+swapping+compression (\texttt{VLLM-SQ}) to KV cache managing, \sys still achieves up to 20$\times$ and on average 9.7$\times$ reduction.

\sys brings substantial context switching speed improvements across various context switching patterns, devices, and LLMs.
We conducted experiments with identical settings for the three patterns (\texttt{Random}, \texttt{Markov} and \texttt{Gaussian}). 
The results indicate that \sys can consistently handle different context access patterns.
On three different devices (Orin/TX2/MI14), \sys consistently demonstrates significant performance improvements.
Notably, on the TX2, the overall switching latency is longer compared to the other two devices. 
This is due to its lower disk bandwidth (SATA HDD) and less-powerful hardware accelerators, which restrict \sys's swapping and swapping-recompute pipeline.
Nevertheless, \sys still outperforms baselines significantly, owing to its context compression and chunk lifecycle management.
Additionally, \sys achieves substantial performance improvements across different LLMs. 
On the OPT model, the switching latency is lower, attributed to its smaller context window (traded for shorter memory and lower in-context learning ability).

\noindent \textbf{Various memory budgets.}
We report the maximal number of active context under different switching latency constraints.
As shown in Figure~\ref{fig:mem_budget}, under a 10 ms latency constraint, \sys supports the switching between 4.32/10.72/16.32 contexts with 1GB/2GB/3GB memory budget, 1.99$\times$/2.48$\times$/2.85$\times$ higher than baselines;
under a 25 ms latency constraint, \sys supports the switching between 9.28/19.34/27.38 contexts with 1GB/2GB/3GB memory budget, 4.16$\times$/3.62$\times$/3.57$\times$ higher than baselines.

\noindent \textbf{Various maximal context lengths.}
We evaluate maximal number of active context under various maximal context lengths.
As shown in Figure~\ref{fig:ctxt_length}, under a 10 ms latency constraint, \sys supports 2.57$\times$/2.95$\times$/3.31$\times$/3.73$\times$/2.24$\times$ more active contexts with 256--4096 maximal context lengths.

\subsection{Compression Efficacy}
\label{subsec: compression}

Recall that \sys employs various levels of compression for different chunks.
We evaluate its overall efficacy and discuss the rationale of parameter selection specified in $\S$\ref{sec:impl}.
In Figure~\ref{fig:quant}, we compare statically quantizing all chunks to the same bitwidth to our quantization.
We choose \cite{xiao2023smoothquant} for static quantization.
The accuracy and compression ratio are averaged across datasets mentioned in Table~\ref{tab:datasets}.
The $ratio_{global}$ and $\{ratio_{w}\}$ are same to $\S$\ref{sec:impl}.

\noindent \textbf{Overall efficacy.}
We observe that our approach outperforms static methods. 
Due to plenty of extreme outliers~\cite{xiao2023smoothquant},  aggressively compressing the entire KV cache into 4-bits/2-bits incurs significant accuracy loss (up to 59\%/99\%). 
In comparison, our method achieves a compression ratio 2$\times$ higher than static methods with negligible loss in accuracy.

\noindent \textbf{Rationale of 4-bits/2-bits compression.}
Generally, as shown in Figure~\ref{fig:quant}, a three levels compression have provided substantial space for almost lossless approximation. 
From the perspective of implementation, 2-bits and 4-bits compression are more hardware-friendly on mobile SoCs.

\subsection{Ablation Study}
\label{subsec: ablation}

We further conduct a breakdown analysis of the benefit brought by \sys’s each technique. 
The experiments are performed on Jetson Orin NX. The results are illustrated in Figure~\ref{fig:ablation}.
We observe that all techniques have non-trivial contribution to the improvement.
For instance, when using Llama2 model to serve 8 active contexts that called in a \texttt{Markov} pattern, \sys takes 0.27 seconds to switch to a new context on average.
Without our chunk lifecycle management, this number becomes 0.62 seconds;
without our tolerance-aware compression or swapping-recompute pipeline, this number becomes 0.42 seconds and 1.62 seconds, respectively.

\subsection{Chunk Size Selection}
\label{sebsec: chunksize}

\begin{figure}[t]
    \centering
    \includegraphics[width=0.42\textwidth]{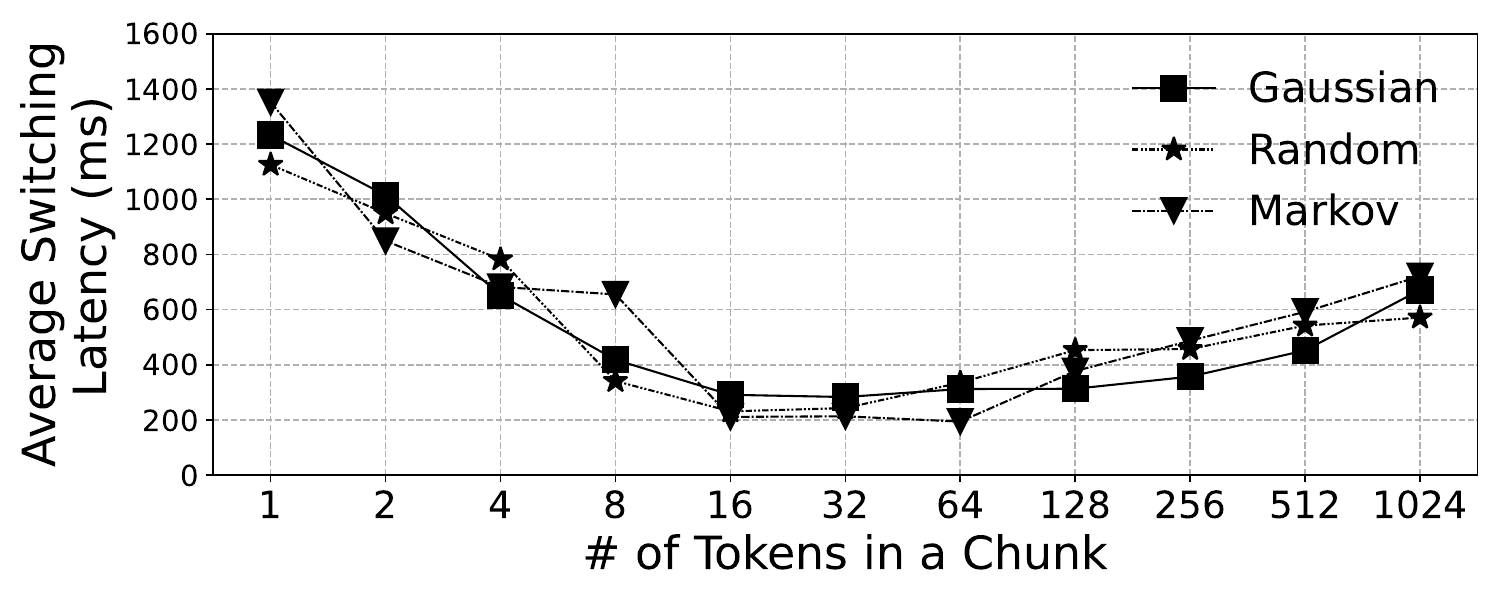}
    \vspace{-12pt}
    \caption{Influence of chunk size. Device: Orin; model: Llama2; active contexts: 8.}
    \label{fig:chunk}
    \vspace{-12pt}
\end{figure}

\sys manages KV cache in chunks.
In our all experiments, the chunk size is set to 16 tokens.
Here we evaluate the influence of chunk size on context switching to validate this setting.
In Figure~\ref{fig:chunk}, we report the 
context switching latency under various token numbers in a chunk.
We observe that a too large or too small chunk size incurs undesirable switching latency.
The reasons are two fold:
small chunks cannot fully utilize disk bandwidth; large chunks result in redundant swapping.
Therefore, \sys selects the optimal trade-off point to fully utilize the idea of chunking.

\subsection{Service Stability Analysis}
\label{subsec: stability}

We analyze \sys's influence on the stability of \service.

\begin{figure}[t]
    \centering
      \begin{minipage}[b]{0.25\textwidth}
        \includegraphics[width=\textwidth]{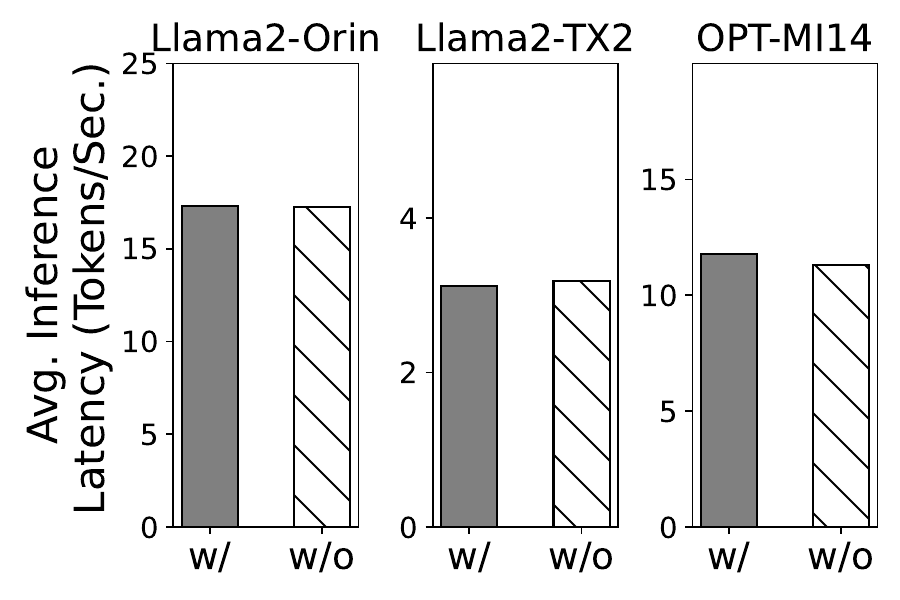}
        \vspace{-17pt}
        \subcaption{Influence on LLM inference performance.}
        \label{fig:influence on inference}
      \end{minipage}
      \begin{minipage}[b]{0.21\textwidth}
        \includegraphics[width=\textwidth]{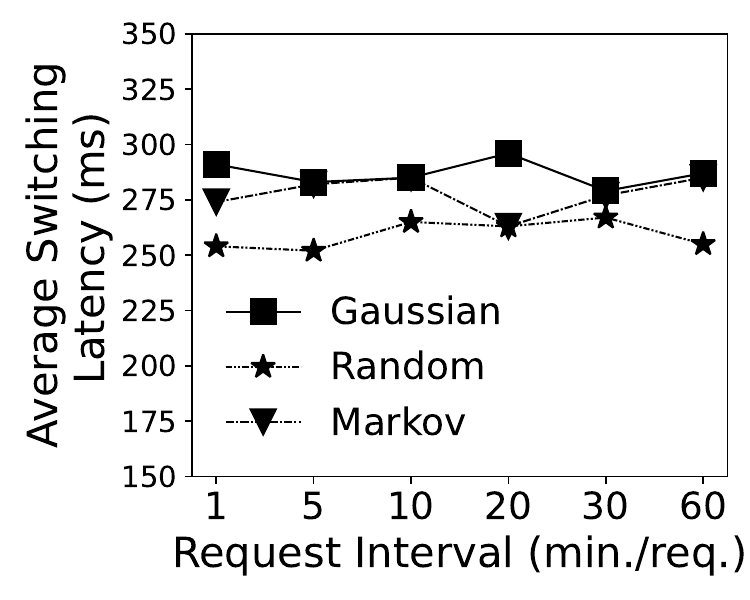}
        \vspace{-17pt}
        \subcaption{Sensitivity to service calling frequency.}
        \label{fig:calling frequency}
      \end{minipage}
    \vspace{-12pt}
    \caption{\sys's influence on LLMaaS stability. ``w/'' and ``w/o'' in (a) means managing context memory with and without \sys, respectively.}
    \label{fig:compression}
    \vspace{-12pt}
\end{figure}

\noindent \textbf{Influence on LLM inference.}
By design, \sys aims to minimize the context switching latency.
In Figure~\ref{fig:influence on inference}, we compare the performance of LLM inference between scenarios with and without the use of \sys.
As observed, there is no significant performance difference (within 5\%).

\noindent \textbf{Sensitivity to service calling frequency.}
Recall that we generate the trace's LLMaaS calling time with a Poisson distribution, which is influenced by the calling rate, i.e., service calling frequency.
In Figure~\ref{fig:calling frequency}, we report the switching latency under various request interval with 16 active contexts on Jetson Orin NX.
\sys demonstrates stable context switching latency at both high and low calling frequencies.

\section{Related Work}


\noindent \textbf{Foundation models on devices.}
AI has become a prevalent workload on mobile devices.
Empowered by the recent progress, especially LLMs in ML community,
some work proposes replacing fragmented task-specific models with a one-size-fits-all foundation model~\cite{yuan2023rethinking, yang2023edgefm, geminiteam2023gemini, touvron2023llama, xu2024survey, liu2023visual, wu2023nextgpt}.
Such models have larger parameter sizes, stronger general-purpose task capabilities, and support multiple modalities.
For instance, M4~\cite{yuan2023rethinking} achieves comparable accuracy to specialized models on five modalities (image/text/audio/IMU/mix) with a multi-path executed architecture.
With the support of corresponding software~\cite{mlc-llm, llama.cpp, song2023powerinfer, mllm, xu2023llmcad} and hardware~\cite{8gen3}, a plethora of revolutionary mobile applications~\cite{xu2024penetrative, glarity, chen2019behavior, customer, news, xiao2023enhancing, wen2023autodroid} are built based on foundation models.
One killer app among them is the LLM agent-based UI automation~\cite{wen2023autodroid, li2024personal_llm_agents}.
For instance, Autodriod~\cite{wen2023autodroid} employs Vicuna~\cite{vicuna2023} to complete an arbitrary task by interacting with the smartphone GUI.
\sys sheds light on deploying the aforementioned foundation models on devices.
At the OS level, \sys provides opportunities to leverage NPU or batching; at the application level, by treating context as LLMaaS interface, \sys can provide personalized and stateful services for apps.

\noindent \textbf{Swapping-based mobile OS memory management.}
A considerable amount of work~\cite{atcswap, drswap, mars, A2S, SmartSwap} employs swapping to mitigate the cold-start latency caused by low-memory killer.
For instance, MARS~\cite{mars} optimizes Linux swapping to enhance performance on flash storage devices. 
By deactivating garbage collection, it reclaims memory from background apps.
Exploiting the opportunity of KV cache, \sys's swapping differs from these approaches.
For instance, its swapping operates at the granularity of token chunks, rather than pages or objects. 
Also, in contrast to lossless app memory compression (e.g., zram~\cite{zram}), \sys introduces approximations for chunks.

\noindent \textbf{KV cache approximation.}
KV cache facilitates LLM in memorizing historical knowledge.
Some recent work~\cite{xiao2023smoothquant, zhang2023h2o, abhyankar2024apiserve, zheng2023efficiently, zaheer2021big} focuses on optimizing KV cache by sparsification or quantization.
Dynamic sparsification methods, such as Big Bird~\cite{zaheer2021big}, only mitigates the compute overhead;
static sparsification methods, such as $H_2O$~\cite{zhang2023h2o}, reduces memory footprint by permanently removing tokens from subsequent LLM decoding.
Regarding quantization, a considerable amount of work~\cite{xiao2023smoothquant, LMDeploy, han2016deep} can losslessly quantize KV cache to 8-bit integers. 
Some work~\cite{zhao2023atom} even propose quantizing it to 4 bits, partly sacrificing generation quality for lower memory consumption.
\sys's compression is orthogonal to these techniques.
Atop them, it further performs more aggressive quantization on less informative chunks to achieve a better accuracy-memory trade-off.
\section{conclusion}
This work advocates LLM as a service on mobile devices (LLMaaS), a new paradigm to fully unleash the power of on-device LLM.
We then present an end-to-end LLMaaS design named \sys with an efficient memory management system of LLM contexts.
\sys enables low-overhead LLM context switching under tight memory constraint through fine-grained, chunk-wise, globally-optimized KV cache compression and swapping.
Extensive experiments demonstrate the efficacy of \sys.

\bibliographystyle{ACM-Reference-Format}
\bibliography{ref-yws}

\end{document}